\documentclass[aps,pre,twocolumn,superscriptaddress,floatfix,10pt]{revtex4}
\usepackage{graphicx}
\usepackage{amsmath}
\usepackage{color}
\usepackage{wrapfig}
\usepackage{latexsym}
\begin{document}

\title{Comparing simulated specific heat of liquid polymers and oligomers to experiments}

\author{Hongyu Gao}
\author{Tobias P. W. Menzel}
\author{Martin H. M\"user}
\affiliation{Dept. of Materials Science and Engineering, Saarland University, 66123 Saarbr\"ucken, Germany}
\author{Debashish Mukherji}
\affiliation{Quantum Matter Institute, University of British Columbia, Vancouver BC V6T 1Z4, Canada}

\begin{abstract}
    The specific heat is a central property of condensed matter systems including polymers and oligomers in their condensed phases. Yet, predictions of this quantity from molecular simulations and successful comparisons to experimental data are scarce if existing at all. One reason for this may be that the internal energy and thus the specific heat cannot be coarse-grained so that they 
    defy their rigorous computation with united-atom models. 
    Moreover, many modes in a polymer barely contribute to the specific heat because of their quantum mechanical nature. Here, we demonstrate that an analysis of the mass-weighted velocity autocorrelation function allows specific heat predictions to be corrected for quantum effects so that agreement with experimental data is on par with predictions of other routinely computed quantities. We outline how to construct corrections for both all-atom and united-atom descriptions of chain molecules. Corrections computed for eleven hydrocarbon oligomers and commodity polymers deviate by less than $k_\textrm{B}/10$ within a subset of nine molecules. Our results may benefit the prediction of heat conductivity. 
\end{abstract}

\maketitle

\section{Introduction} 
\label{sec:introduction}
The molecular simulation of polymers and oligomers is in a mature state, which allows chemistry-specific predictions of many physical properties to be made.
This includes, in particular, the prediction, or, reproduction, of density~\cite{Vagilis11mac,Du12JCTC}, viscosity~\cite{Kremer98BI,Bair2002PRL,Habchi2010TI,Jadhao2019TL}, and mechanical properties~\cite{Goddard91JPC,Rutlege94JPC,Root16Mac} as functions of temperature, pressure, and shear rate but also the computation of complex phase diagrams~\cite{Kroeger04PR,Mueller20PPS,Mukherji20AR}. 
Molecular simulation has even reached levels making it possible to design lubricants with small viscosity index~\cite{Kajita2020CP}.
However, we did not manage to find any successful predictions for the specific heat $c_p$ of systems containing chain molecules, although, in principle, the specific heat could (falsely) be deemed a profane property to compute.
It only requires the temperature derivative of the enthalpy to be taken and/or the energy or enthalpy fluctuation to be determined. 
There are certainly two main reasons impeding the calculation of the specific heat from molecular simulations.
First, united-atom descriptions ignore the presence of hydrogen atoms so that their small but non-zero contribution to $c_p$ is ignored.
Second, and more importantly, both united-atom and all-atom descriptions generally assume nuclei to be classical objects, while in reality, their motion is quantum mechanical.
This difference makes classical simulations overestimate the specific heat at small temperatures.
It explains why Bhowmik \textit{et al.}~\cite{Bhowmik2019B} found that the heat predicted from classical all-atom molecular dynamics (MD) simulations of hydrocarbon chains was almost a factor of three too high, while results for polytetrafluoroethylene (PTFE) exceeded experimental values only by 20\%.

These findings can be rationalized in a back-of-the-envelope calculation.
The vibrational energy of a CF bond is near 20~THz while that of the CH bond lies near 90~THz.
At room temperature, each such mode contributes to the specific heat with approximately 0.45~$k_\textrm{B}$ (CF) and $1\cdot 10^{-4}$~$k_\textrm{B}$ (CH), respectively, while a classical harmonic mode would contribute $k_\textrm{B}$ according to the Dulong-Petit law. 
Many other modes also become more classical in PTFE compared to hydrocarbon chains, because fluorine atoms are heavier than hydrogen atoms, while bond stiffnesses do not depend substantially on the termination. 
Approximating all modes in PTFE other than the CF-stretching bond as perfectly classical would suggest that a classical PTFE simulation at room temperature should be reduced by twice 0.55~$k_\textrm{B}$
per CF$_2$ repeat unit, so that the quantum effect of the CF vibration can be estimated to reduce the specific heat of PTFE by roughly 15\%. 
A similarly accurate estimate for hydrocarbons is difficult to make, because a rather large fraction of characteristic frequencies require corrections spanning the entire domain from very small to unity.
However, for a crude approximation, one could argue hydrogen atoms to be completely quantum and carbon atoms to be close to classical.

One possibility to account accurately for the quantum nature of nuclear degrees of freedom is to treat them in a path-integral framework, as done more than 20 years ago by Marto{\v{n}}{\'{a}}k \textit{et al.}~\cite{Martonak1998PRE}. 
However, this approach is computationally  demanding.
Reaching the proper quantum limit needed for a reasonably accurate, direct estimate for condensed matter systems necessitates the simulations of $P$ replica of the system, where the so-called Trotter number $P$ needs to slightly exceed the ratio $h\nu/k_\textrm{B}T$~\cite{Muser1995PRB,Herrero2014JPCM}.
Here $h$ is Planck's constant, $\nu$ is the maximum characteristic frequency in the system (e.g., the CH bond-stretching vibration), while $k_\textrm{B}T$ is the thermal energy. 
A related approach to simulate the effect of quantum mechanics is the use of potentials that implicitly include quantum effects through the Wigner-Kirkwood expansion~\cite{Wigner1932PR,Kirkwood1933PR} of the free energy in powers of Planck's constant. 
Using the leading-order terms, the temperature range, in which experimental data on the specific heat of magnesium oxide was successfully reproduced, extended to temperatures a little below the Debye temperature, but not further below~\cite{Matsui1989JCP}.
Moreover, both the extra programming and computing time associated with the Kirkwood-Wigner expansion exceed that by path integrals substantially, so that an alternative, feasible, and easy-to-implement way to correct the specific heat of polymeric systems for quantum effects remains sought after.  

In this paper, we extend a method introduced by Horbach \textit{et al.}~\cite{Horbach1999JPCB} to calculate the low-temperature specific heat of a quantum mechanical system, namely silica well below its glass transition temperature.
To this end, they first computed the mass-weighted, velocity autocorrelation function $C(\Delta t)$ using classical MD.
For a fictitious harmonic reference yielding the same $C(\Delta t)$, the Fourier transform of this function, $g(\nu)$, allows the vibrational density of states to be directly deduced and from it the specific heat. 
Rather than to report that number directly, as done by Horbach \textit{et al.}~\cite{Horbach1999JPCB}, we use it to estimate the specific-heat \emph{difference} between a classical system and a corresponding quantum mechanical system.
This way, we correct predominantly the stiff, high-frequency modes, which should obey the harmonic approximation reasonably well, while leaving the specific-heat contributions of the slow modes unaffected.
The latter are certainly anharmonic in the liquid phase, whereby they contribute in a non-trivial fashion to the heat balance. 

Specific heats obtained in simulations not containing all degrees of freedom (DOFs) explicitly, such as in coarse-grained models,
cannot be corrected as straightforwardly as those measured in classical all-atom simulations representing all DOFs explicitly. 
The optimum way to proceed depends not only on the type of coarse graining but also on whether or not an (unconstrained) all-atom simulation can be conducted at one or two representative temperatures.
Thus, several avenues to estimate specific-heat corrections due to missing hydrogen atoms will also be discussed in this work. 

The remainder of this article is organized as follows:
the simulation methods are presented in Sect.~\ref{sec:method}.
Sect.~\ref{sec:theory} describes our approach to correcting specific heats.
Sect.~\ref{sec:results} contains the results.
Conclusions are drawn in Sect.~\ref{sec:conclusions}. 

\section{Simulation Methods}
\label{sec:method}

The simulations in this work were conducted by three different people, each one with his own preferences for software, potentials, and other details pertaining to methods, such as thermostats.
Since all of the choices are made routinely in different contexts, the diversity of approaches allows the robustness of the observed trends to be tested. 

\begin{figure}[ptb]
\includegraphics[width=0.49\textwidth,angle=0]{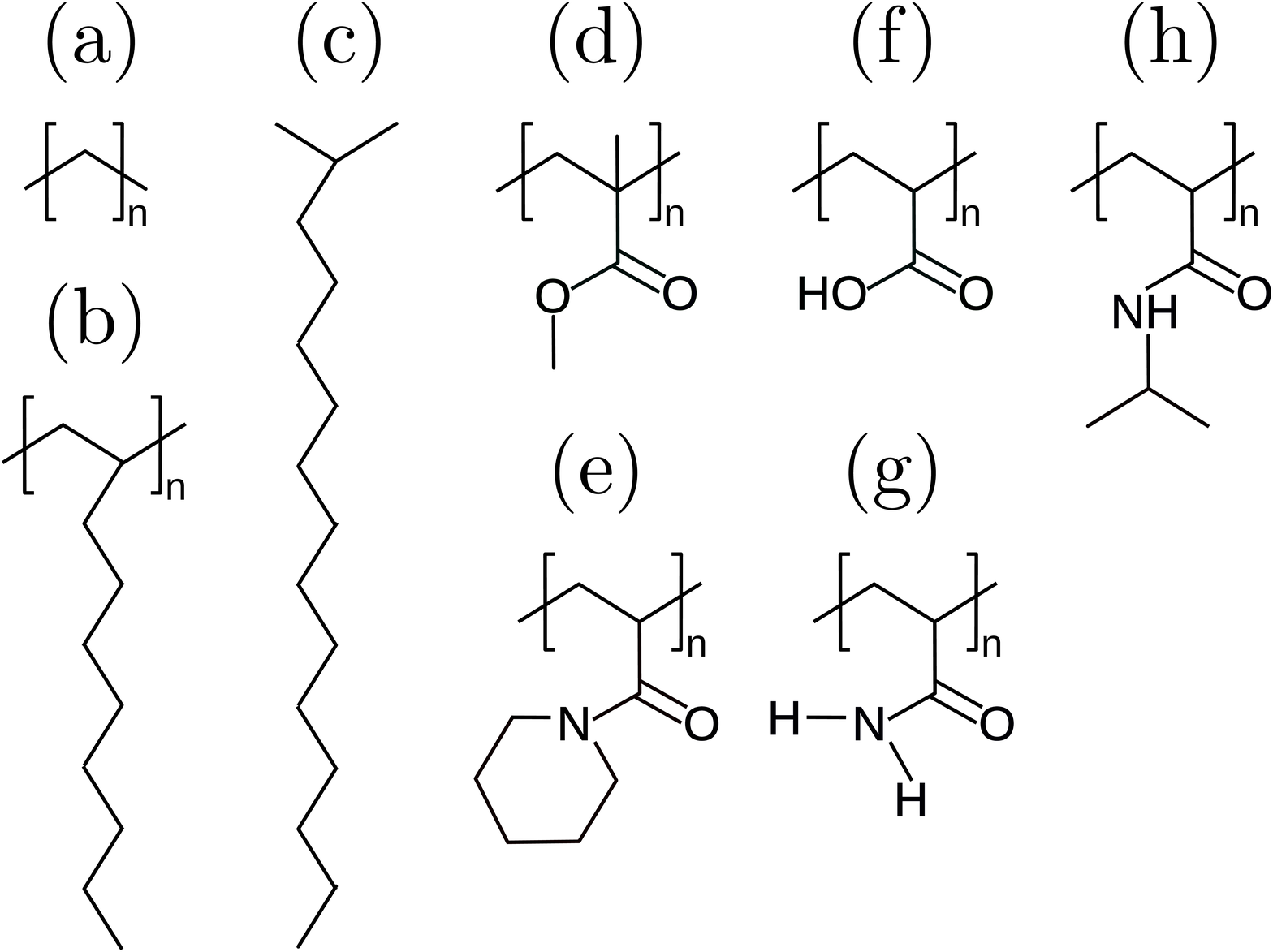}
        \caption{Schematics showing different monomeric structures investigated in this study. Parts (a-c) show hydrocarbon structures for  $n$-octane ($n=8$, including end groups) and $n$-hexadecane ($n=16$) in part (a), decene-dimer ($n=2$), -trimer ($n=3$) and -tetramer ($n=4$) in part (b), and isohexadecane in part (c). Parts (d-h) show commodity polymer structures for poly(methyl methacrylate) (PMMA), poly(N-acryloyl piperidine) (PAP), poly(acrylic acid) (PAA), poly(acrylamide) (PAM), and poly(N-isopropyl acrylamide) (PNIPAM), respectively. Note that for parts (a-b) and (d-h) chain ends outside the bracket are terminated with hydrogen atoms.
        \label{fig:schem}}
\end{figure}

For this study we chose two different sets of chain molecules: (1) linear and branched hydrocarbon oligomers and (2) commodity polymers containing elements in addition to carbon and hydrogen in the repeat units, see Fig.~\ref{fig:schem} for more details of the molecular structures. All simulations were conducted in the $NpT$-ensemble at atmospheric pressure.
Temperature $T$ was raised from $T \approx 300$~K to $T \approx 560$~K for all hydrocarbons, except for $n-$octane for which $T$ varied from 200~K to 380~K. In the case of commodity polymers, $T$ lied between $440-600$~K. 

The specific heat was computed in two ways:
first, by taking finite differences of the enthalpy $H(T)$ according to
\begin{equation}
\label{eq:finiteDiffcP}
    c^\textrm{cla}_p(T) \approx \frac{H(T+\Delta T) - H(T-\Delta T)}{2\,\Delta T},
\end{equation}
and second by fitting a third-order polynomial to $H(T)$.
Since the temperature-dependence of $c_{p}$ is rather weak in the considered temperature range, the second method may be slightly preferable. 

For the initial set of hydrocarbon simulations, we have chosen six different linear and branched oligomers, see Figs.~\ref{fig:schem}(a-c). The all-atom simulations are performed using the LAMMPS molecular dynamics package~\cite{Plimpton1995JCP}. The improved L-OPLS-AA force field parameters are used to simulate the all hydrocarbons~\cite{Price2001JCC,Siu2012JCTC}, except for $n$-octane where we have used the standard OPLS-AA \cite{OPLS}. 
The potentials were chosen because they reproduced experimental data on density, viscosity, and diffusion coefficient quite accurately~\cite{Siu2012JCTC}.

The number of chains in a cubic simulation box was adjusted such that each system consists of approximately $10^4$ atoms. The temperature and pressure are imposed using the Nos${\rm\acute e}$-Hoover thermostat and barostat, respectively. For the  temperature coupling, the time constant is chosen as $\tau_{T} = 0.1$~ps and for pressure as $\tau_{p} = 1$~ps. The long-range electrostatic interactions are treated using the particle-particle particle-mesh (PPPM) solver \cite{Hockney1988}. The interaction cutoff is chosen as $r_c = 1$~nm. The simulations for $n$-octane and $n$-hexadecane was performed for 6 ns, while for the other hydrocarbon oligomers we have conducted 10 ns simulations. These simulation time scale ensure well equilibration of the samples and the average of $H(T)$ is calculated by taking the last 2 ns data. The typical time step for the all-atom simulation is chosen as $\Delta t = 1$~fs. 
 
For $n$-hexadecane, we have also performed simulations using the united-atom TraPPE-UA force field~\cite{Martin1998JPCB}. In this case, the employed time step was set to $\Delta t = 2$~fs.

For the second set of systems, we investigated five different commodity polymers, namely poly(methyl methacrylate) (PMMA), poly(N-acryloyl piperidine) (PAP), poly(acrylic acid) (PAA), poly(acrylamide) (PAM), and poly(N-isopropyl acrylamide) (PNIPAM), see Figs.~\ref{fig:schem}(d-h). The choice of these polymers is motivated by their possible use for the design of advanced polymeric materials~\cite{Cahill16Mac,Mukherji19PRM}. The chain length $N = 30$ is taken for PMMA, PAP, PAA and PAM, while $N = 40$ for PNIPAM. 
Different number of repeat units were used, because all-atom chain configurations were available from  earlier studies by one of us \cite{Mukherji19PRM,Mukherji17NC,Mukherji17JCP}. Each configuration consists of 100 polymer chains randomly distributed within a cubic simulation box. All these polymers were equilibrated earlier in their (solvent free) melt states at T = 600~K, which is at least 150~K above their calculated glass transition temperatures~\cite{Mukherji19PRM}.

All commodity polymers are modelled only in the full atomistic description. The standard OPLS-AA force field parameters \cite{OPLS} are used for PAP, PAA, and PNIPAM, while the modified parameters are used for PMMA \cite{Mukherji17NC} and PAM \cite{Mukherji17JCP}.
The used potential reproduce not only bulk polymer properties, such as the density and elastic response \cite{Mukherji19PRM}, but also capture their solvation in dilute aqueous solutions~\cite{Mukherji17NC,Mukherji17JCP}.

The simulations of commodity polymers are performed using the GROMACS molecular dynamics package \cite{gro}. 500~ns long, $NpT$ simulation were conducted for each system at each temperature. The total accumulated MD time for the commodity polymers is 25 $\mu$s. Here, the temperature is imposed using the ``canonical-sampling-through-velocity-rescaling thermostat''~\cite{Vscale} with $\tau_{T}=1$~ps  and the pressure is set to 1~atm with a Berendsen barostat using $\tau_p = 0.5$~ps \cite{Berend}.
Electrostatics are treated using the particle-mesh Ewald method \cite{PME}. The interaction cutoff for non-bonded interactions is chosen as 1.0 nm. The simulation time step is taken as $\triangle t = 1$ fs and the equations of motion are integrated using the leap-frog algorithm. For the calculation of $H(T)$, we have used the last 50~ns data after $H(T)$ reached a reasonable plateau.

All polymeric systems described above are simulated in their liquid phase, where the equilibration of the individual samples are still possible. Moreover, for the case of $n$-octane we have also performed simulations with a crystalline phase at $T = 40$ K and a quenched phase, where a $n$-octane liquid at $T = 300$ K was shock-quenched to $T = 40$ K.

\section{Theory} 
\label{sec:theory}

\begin{figure}[hbtp]
    \centering
    \includegraphics[width=.4\textwidth]{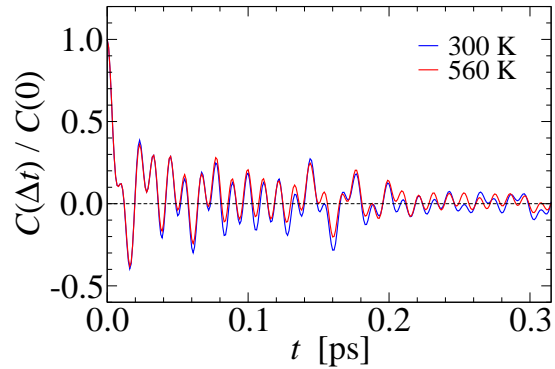}
    \caption{Normalized mass-weighted velocity autocorrelation function $C(\Delta t)/C(0)$ of hexadecane at temperature $T = 300$~K (blue) and at $T = 560$~K (red). 
    }
\label{fig:hexadecaneACF}
\end{figure}

The central property to be computed in this work is the mass-weighted velocity autocorrelation function (ACF),
\begin{equation} \label{eq:globalACF}
    C(\Delta t) = \sum_n m_n \, \left\langle \mathbf{v}_n(t) \cdot \mathbf{v}_n(t+\Delta t) \right\rangle,
\end{equation}
where $m_n$ is the mass of atom $n$ and $\mathbf{v}_n(t)$ its velocity at time $t$, while the angles
$\langle ... \rangle$ denote a thermal equilibrium average.
A typical example for $C(\Delta t)$ is presented in Fig.~\ref{fig:hexadecaneACF}.
It shows long-lived fluctuations, unlike the velocity ACF of simple liquids, in which all interactions are of similar strength.

Depending on whether $C(\Delta t)$ is measured using the coarse-grained descriptions of the polymer, as for the united-atom potentials, or in an all-atom simulation, different strategies can be pursued to estimate how the specific heat needs to be corrected to account for nuclear quantum effects.
These are described in the following. 
 
\subsection{All-atom descriptions}

In an equilibrated harmonic system, as much energy is contained in the potential energy as in the kinetic energy.
If the frequency of a harmonic mode is known, e.g., from the measurement of its classical velocity ACF, the specific heat of this mode after quantization is given by
\begin{equation}
    c_p^\textrm{qm}(\nu,T) = k_\textrm{B} \frac{(h\nu/2k_\textrm{B}T)^2}{\sinh^2(h\nu/2k_\textrm{B}T)},
\end{equation}
as can be easily derived from the partition function of the quantum mechanical harmonic oscillator, see Ref.~\cite{Horbach1999JPCB}, or most textbooks on statistical mechanics.
Since the specific heat of a classical harmonic mode satifies the Dulong-Petit law, $c_p^\textrm{cla}(T) = k_\textrm{B}$, the difference between the specific heat of a classical and a quantum system simply is $\Delta c_p = k_\textrm{B} - c_p^\textrm{qm}$ for each degree of freedom (DOF). 

In a harmonic system, the global ACF defined in Eq.~(\ref{eq:globalACF}) results from the superposition of individual normal modes so that its Fourier transform allows us to determine what percentage of modes has what resonance frequency.
Towards this end, we define the spectrum
\begin{equation}
\label{eq:DOSfromACF}
    g(\nu) =  
    \frac{1}{G} \int_0^\infty \!\! \mathrm{d}t\,
    \cos(2\pi\nu\Delta t) \,
    \frac{C(\Delta t)}{C(0 )},
\end{equation}
where we have divided $C(\Delta t)$ by 
$C(0)$, whose exact value is 
$D\,N\,k_\textrm{B}T$, where $D=3$ is the spatial dimension and $N$ the number of explicitly considered atoms. 
Finally, we chose the prefactor $G$ in Eq.~(\ref{eq:DOSfromACF}) such that the integral over $g(\nu)$ is unity.
This way, $g(\nu)$ can be interpreted as the vibrational density of states (DOS) normalized to an individual degree of freedom and in a unit system, in which Planck's constant defines the unit of angular momentum. 
The typical DOS for all molecules in Fig. \ref{fig:schem} are shown in the Supplementary Fig. S1~\cite{SM}.

The relative difference between the specific heat of a classical and a quantum system can now be obtained as
\begin{equation} \label{eq:specificHeat}
    \Delta c_\textrm{rel}(T) = \int_0^\infty \!\! \mathrm{d}\nu\,
    g(\nu)\,\left\{ 1 - c_p^\textrm{qm}(\nu,T)/k_\textrm{B}\right\}.
\end{equation}
Thus, the specific heat of a system of quantum mechanical harmonic oscillators would read 
\begin{equation} \label{eq:correctAllAtom}
    c_p(T) = c_p^\textrm{cla}(T)-c_p^\textrm{DP}\, \Delta c_\textrm{rel}(T),
\end{equation}
where $c_p^\textrm{cla}(T)$ is the specific heat of the classical system and 
$c_p^\textrm{DP}$ the specific heat of the system assuming the Dulong-Petit law to be valid, i.e.,
$c_p^\textrm{DP} = k_\textrm{B} n_\textrm{DOF}$, where $n_\textrm{DOF}$ is the number of DOFs. 

We propose to use Eq.~(\ref{eq:correctAllAtom}) for any system, whose degrees of freedom can be partitioned into slow modes, which are typically soft and/or anharmonic, and high-frequency modes, which tend to be quasi harmonic. 
This procedure leaves (low-frequency) contributions to the specific heat that deviate from Dulong-Petit's law unchanged, but distinctly reduces the specific heat associated with the high-frequency modes involving hydrogen atoms. 

Ideally, $g(\nu)$ is determined in the vicinity of the temperature at which the specific heat is computed.
However, we demonstrate in Sect.~\ref{sec:results} that the high-frequency spectra and thereby the specific-heat corrections are relatively insensitive to the temperature at which $g(\nu)$ is determined.
Thus, it should be generally sufficient to compute $g(\nu)$ at a single, medium temperature, or, alternatively to compute $g(\nu)$ at the lowest and highest temperature and to interpolate continuously between the spectra (or the two subsequent specific heat corrections) at intermediate temperatures.

\subsection{United-atom descriptions}
\label{sec:methodB}
In united-atom descriptions and/or when using bond length constraints, the number of degrees of freedom (DOFs) is reduced compared to the real system.
While only stiff modes not contributing significantly to the specific heat are usually eliminated in chemistry-specific,  coarse-grained descriptions of polymers, a precise calculation of $c_p$ may necessitate the estimation of the contribution of the eliminated DOFs to the specific heat. 
Thus, the full (quantum) contributions of the $N_\textrm{ig}$ ignored degrees of freedom to $c_p(T)$ must be added to the estimate of the $N_\textrm{ex}$ explicitly treated degrees of freedom.
If specific heats are normalized to individual degrees of freedom, this yields
\begin{equation} \label{eq:fullUA-cp}
    c_p(T) = \frac{N_\textrm{ex}\,c_p^\textrm{ex}(T)\,\{1 -\Delta c_\textrm{rel}^\textrm{ex}(T)/k_{\rm B}\}+N_\textrm{ig}\,c_p^\textrm{ig}(T)}{N_\textrm{ex}+N_\textrm{ig}},
\end{equation}
where the contribution of the ignored DOFs can be estimated with the help of the density of states associated with the motion of the ignored DOFs, $g_\textrm{ig}(\nu)$, i.e., with 
\begin{equation}
c_p^\textrm{ig}(T) = 
 \int_0^\infty \!\! \mathrm{d}\nu\,
    g_\textrm{ig}(\nu)\,c_p^\textrm{qm}(\nu,T).
\end{equation}
In the following, we propose three different ways to estimate the density of states of the ignored degrees of freedom.

\subsubsection{Difference method}

In the first method, which we call the difference method, the all-atom and the united-atom $g(\nu)$ are both computed and normalized to the same entity, e.g., to a single polymer or to an atom as in a count of all atoms, including those that were eliminated in the united-atom simulation. 
The missing contribution then reads $g_\textrm{ig}(\nu) = g_\textrm{aa}(\nu) - g_\textrm{ua}(\nu)$.
Note that $g_\textrm{ig}(\nu)$ may have negative contributions, which, however do not cause any trouble in practice.

\subsubsection{Explicit method}

In the second method, which we call the explicit method, an all-atom system is first equilibrated at a representative temperature.
All heavy atoms are then fixed in space and only hydrogen atoms are propagated in time and thermostatted, however, only so moderately that peaks in $g(\nu)$ do not broaden substantially. 
In this follow-up simulation, the hydrogen velocity ACF is measured and a first estimate for $g_\textrm{ig}(\nu)$ is obtained through a Fourier transform of that ACF.
Since the mass of carbon atoms is finite, we suggest to reinterpret a frequency $\nu$ as $\alpha\nu$ with $\alpha=\sqrt{13/12}$ so that reduced-mass effects are  accounted for approximately. 
At the same time, it needs to be ensured that the integral over $g_\textrm{ig}(\nu)$ yields the relative number of hydrogen atoms so that the full transformation can be cast as $g(\nu)\to g(\alpha\nu)/\alpha$.

\subsubsection{Crude method}

While only one or at most two all-atom simulation need to be run for the difference method and the explicit method to be executed, it might still be beneficial if setting up an all-atom system can be avoided all together. 
We thus need a third way to compute specific heat corrections, which could be called the I-don't-want-to-run-an-all-atom-simulation-but-still-need-a-rough-guess-for-the-specific-heat-correction method (quantum chemists would probably introduce the catchy and easy-to-remember abbreviation IDW2RA3SBSNARG4TSHC).
To this end, we suggest to approximate $g_\textrm{ig}(\nu)$ with a set of delta-functions:
\begin{equation}
    \label{eq:IDW2a}
    g_\textrm{ig}(\nu) = n_\textrm{H in CH$_x$}^\textrm{rel}\,\sum_{i=1}^{n_x} w_{x,i} \,\delta(\nu-\nu_{x,i}),
\end{equation}
where $n_\textrm{H in CH$_x$}$ with $x = 2$ or $3$ is the relative number of hydrogen atoms being part of a CH$_2$ or CH$_3$ unit, respectively, while the $w_{x,i}$ are weights and the $\nu_{x,i}$ are frequencies.
We describe in the Supplementary Information how the pairs $(w_{x,i}, \nu_{x,i})$ were obtained and merely note their results here.
For CH$_2$, we used $(1/6,20)$, $(1/2, 37.5)$, and $(1/3, 90)$. 
For CH$_3$, we used $(1/9, 8.5)$, $(1/9, 23)$, $(1/9, 30)$, $(2/9, 39)$, $(1/9, 50)$, $(1/9, 75)$, and $(2/9, 93)$.
Frequencies are stated in THz.

\subsubsection{Comparison of united-atom correction methods}

The difference method is directly applicable to coarse-graining approaches going beyond the elimination of hydrogen atoms.
The same holds for the explicit method, however, with the constriction that the corrective factor $\alpha$ would have to be modified when deuterium atoms are involved and/or hydrogen atoms terminate other atoms than carbon atoms. 
The crude method is only meant to be used directly when hydrogen atoms bonded to carbons are eliminated. 
When all hydrogen terminations are replaced with deuterium atoms, it might suffice to divide all used frequencies with $\sqrt{2}$.
However, simple rescaling of frequencies would not be advised for partial deuterium termination.

Finally, we note that a highly accurate knowledge of the respective spectra is not needed, unless $c_p$ must be known with a great accuracy. 
If a vibrational frequency has an error of say 10\%, which most contemporary force-fields should be in a position to reproduce, then the temperature range in which the absolute error of the quantum correction exceeds 0.1~$k_\textrm{B}$ of that mode is roughly $0.3 < k_\textrm{B}T/(h\nu) < 1.2$.
Since the density of states spans a broad range of frequencies, the relative number of modes lying in such a range is typically at best around 30\%.

\section{Results} \label{sec:results}

\subsection{Explicit-atom simulations}

\begin{figure*}[ptb]
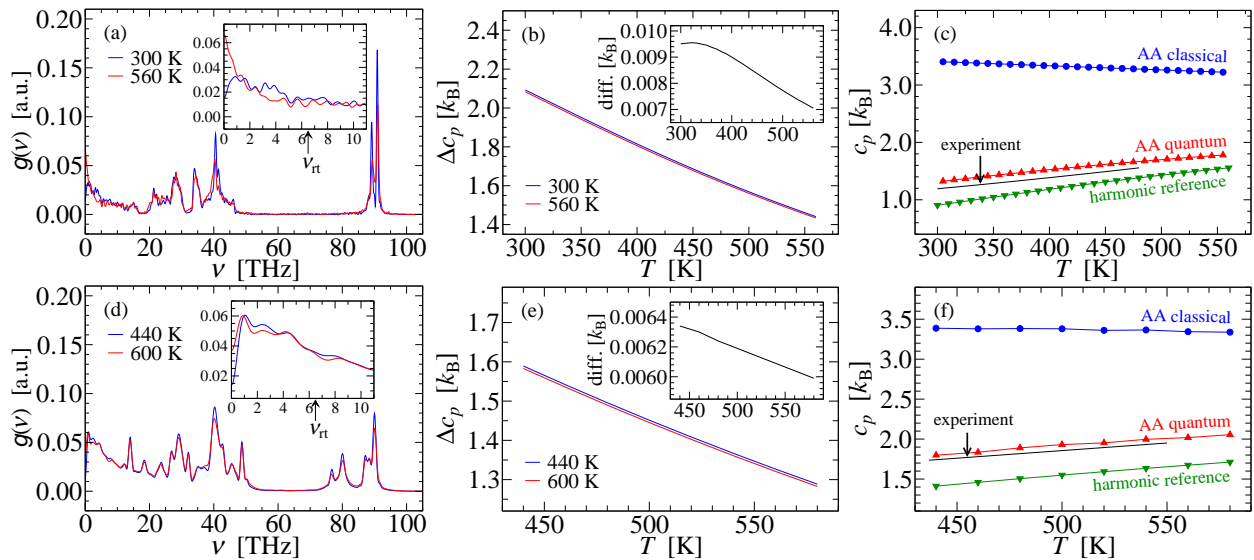

    \centering
    \includegraphics[width=.309\textwidth]{spectrum_hexadecane_aa.eps}
    \includegraphics[width=.3\textwidth]{delta_cp_hexadecane_aa.eps}
    \includegraphics[width=.3\textwidth]{cp_hexadecane_aa.eps}
    \includegraphics[width=0.309\textwidth]{spectrum_pmma_aa.eps}
    \includegraphics[width=0.3\textwidth]{delta_cp_pmma_aa.eps}
    \includegraphics[width=0.3\textwidth]{cp_pmma_aa.eps}
 \caption{ \label{fig:HD_PMMA} Vibrational spectra $g(\nu)$, panels (a) and (d), specific heat corrections, $\Delta c_p$, (b) and (e), as well as specific heats, (c) and (f), for hexadecane in the top row (a)--(c), and for PMMA in the bottom row (d)--(f).
 In each case, $g(\nu)$ was obtained at a low (blue) and a high (red) temperature and $\Delta c_p$ deduced from it. The corresponding blue and red curves essentially overlap in panels (b) and (e). 
 Their differences (diff.) are shown in their insets. 
 Experimental data on $c_p$ for $n-$hexadecane~\cite{Regueira2017JCT} and PMMA~\cite{expPMMA} are shown in black lines.
 They are compared to three numerical data sets:
 classical all-atom simulations (blue circles), results obtained using the harmonic~reference method~\cite{Horbach1999JPCB} (green triangles down) and from the methodology proposed in this work (red triangles up).
 }
\end{figure*}

The first step of estimating the specific-heat corrections in an explicit-atom simulation consists of measuring the full mass-weighted velocity ACF, $C(\Delta t)$, which is worth discussing in its own right.
Fig.~\ref{fig:hexadecaneACF}  shows $C(\Delta t)$ for $n$-hexadecane at the lowest and highest temperature investigated, i.e., at $T = 300$~K and at $T = 560$~K, each time normalized such that $C(0) = 1$.
Both correlation functions have maxima and minima at similar locations.
Peak heights and intensities are almost identical at very small times but start to differ at large times. 
As a consequence, the Fourier transform of $C(\Delta t)$, a.k.a. spectra or DOS's, which is shown in Fig.~\ref{fig:HD_PMMA}(a), is essentially identical at high frequencies for $300$ and $560$~K. 
Significant differences appear only at frequencies below what could be called the thermal frequency, which we define as $\nu_\textrm{t}=k_\textrm{B}T/h$.
The numerical value of the ``room-temperature thermal frequency" is $\nu_\textrm{rt} = k_\textrm{B}~300~\mathrm{K}/h \approx 6.25$~THz.

Since the $c_p$ correction for a single mode with thermal frequency is merely around 8\%, the total specific heat corrections are rather insensitive to the temperature, at which the DOS was deduced, as long as that temperature lies in a reasonable interval.
This claim is confirmed in panel (b) of Fig.~\ref{fig:HD_PMMA}, particularly in its inset, where differences between the $c_p$ corrections obtained at 300 and 560~K are shown to differ by no more than 0.5\%.

Fig.~\ref{fig:HD_PMMA}(c) confirms the previously made observation~\cite{Bhowmik2019B} that classical, all-atom-based simulations of chain molecules with hydrogen termination overestimate the specific heat at room temperature by a factor smaller than but close to three. 
The discrepancy reduces with increasing temperature, but is still close to a factor of two at $T = 550$~K. 
However, after applying the specific-heat corrections to the classical $c_p(T)$ data, agreement with experimental results is obtained within 0.1~$k_\textrm{B}$ per atom, which translates to a relative accuracy of approximately 6\%. 
At the same time, our analysis reveals that the specific heat of the harmonic reference is clearly below both experimental data sets.
Thus, while the original correction method pursued by Horbach \textit{et al.}~\cite{Horbach1999JPCB} clearly reduces the error from approximately 200\% to 20\%, our modification reduces the error by another factor of three. 
We note in passing that our treatment would not have improved the accuracy of the $c_p$ prediction for their system in a similar fashion, as they kept their supercooled silica at a relatively small temperature, where thermal anharmonicity effects are small. 

The just-reported methodology was repeated for all investigated systems.
However, only one more example is presented explicitly, namely PMMA in Fig.~\ref{fig:HD_PMMA}(d)--(f).
At high frequencies, an additional (double) peak shows up in $g(\nu)$ near 80~THz, which we attribute to the H vibrations of the methyl group attached to the side group, while the extra peak at 50~THz is due to the stretching vibrations of the CO double bond.
Differences between spectra measured at different temperatures are again only substantial at frequencies at or below the lower of the two investigated temperatures, this time $T = 440$ and 600~K. 
Thus, specific heat corrections are again essentially identical irrespective of the temperature at which the DOS was acquired.
Finally, Fig.~\ref{fig:HD_PMMA}(f) confirms that the original harmonic reference reduces the $c_p$ deviation between classical simulations of hydrocarbons and experiment by a factor close to ten and that using the proposed difference-methodology reduces the error much further.
Given the currently available data, agreement appears to be within 2\%. 

At this point, it is difficult to speculate what the main reason for the small \emph{absolute} discrepancies between experimentally and \textit{in-silico} measured specific heats of order 0.1~$k_\textrm{B}$ may be, i.e., if they are mainly due to errors in the classical reference or if they originate from the quantum corrections, or, unlikely but not impossible, if they stem from experimental errors. 
Irrespective of the answer to this question, it appears to us that simulations should be in a position to predict specific heat \emph{differences} between different polymers to within clearly less than 0.1~$k_\textrm{B}$, at least as long as consistent potentials are used, i.e., it should be ensured that dispersive interactions, bond stiffnesses, bond angles, etc. are parameterized consistently when trying to ascertain specific heat differences between two liquids.
This way, absolute errors would be highly correlated so that differences between the specific heat of different liquids can be resolved with great accuracy.  

\begin{figure}[hbtp]
    \centering
    \includegraphics[width=.4\textwidth]{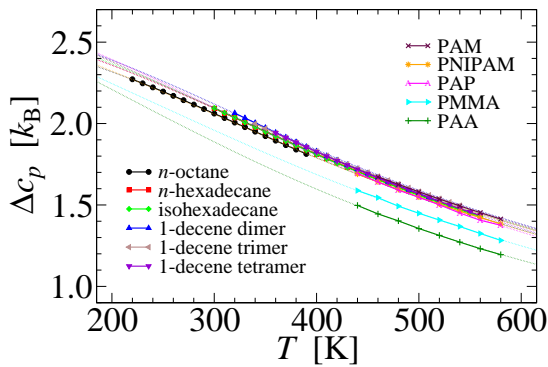}
    \caption{Specific heat corrections for explicit-atom simulations of all chain molecules investigated in this study. Only lines are shown at temperature, where commodity polymers could not be equilibrated using feasible computing times.
    }
    \label{fig:cpCorr}
\end{figure}

An interesting observation that can be made when comparing the simulation data for hexadecane (HEX) and PMMA is that the specific-heat corrections at 450~K are quite similar, i.e., 1.68~$k_\textrm{B}$ (HEX) versus 1.56~$k_\textrm{B}$ (PMMA).
In fact, Fig.~\ref{fig:cpCorr} reveals that the specific heat correction of most of the investigated molecules obey an almost universal function $\Delta c_p(T)$ in the investigated temperature range within less than 0.1~$k_\textrm{B}$.
However, even the two exceptions, namely PMMA and PAA, do not stray too far away from the general trend.
This is somewhat surprising given the significant differences in the monomer architectures shown in Fig.~\ref{fig:schem}.
The relatively small $\Delta c_p$ of PAA can be rationalized as follows:
The side group provides an extra classical degree of freedom, i.e., the libration of the side group, while having only one hydrogen atom per three heavy atoms. 
The $g(\nu)$, from which the $c_p$ corrections presented in Fig.~\ref{fig:cpCorr} were deduced, are shown in the Supplementary Figs. S1~\cite{SM}.

Unfortunately, we did not manage to improve the superposition of the various $\Delta c_p(T)$ curves by scaling the corrections with the relative (inverse) ratio of estimated ``quantum'' DOFs per total DOFs. 
Thus, at this point of time, we can only recommend to use the quasi-universal correction for those (carbon-based molecules with predominant hydrogen termination) polymers  that are not included in our list, for a ``quick and dirty'' assessment of the specific heat from classical explicit-atom simulations.

The $c_p$ corrections do not appear to change substantially upon crystallization.
For octane we found $\Delta c_p$ estimated from a 40~K crystal to exceed that deduced from a 300~K liquid, both at atmospheric pressure, by approximately 0.05~$k_\textrm{B}$ per DOF in between these two limits, see the Supplementary Figs. S2(b) and (d)~\cite{SM}.
The increase is predominantly due to the fact that the ordering and the subsequent densification of octane increases vibrational frequencies, because atoms are pushed more deeply into the stiff, repulsive part of their interaction.  
A similar comment holds for pressurized liquids when setting the pressure in a $n-$octane at 2 and 4~GPa. The corresponding data is shown in the Supplementary Figs. S2(a) and (c)~\cite{SM}. 

\begin{figure}[hbtp]
\centering
\includegraphics[width=.4\textwidth]{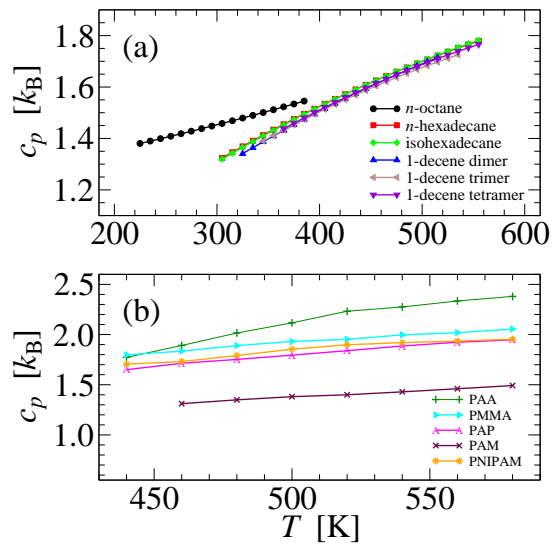}
\caption{Specific heat predictions from all-atom simulations of various chain molecules after applying quantum corrections grouped into (a) hydrocarbon oligomers and (b) commodity polymers.}
    \label{fig:cpall}
\end{figure}

Of course, it is only worth knowing $\Delta c_p$ if variations in $\Delta c_p$ from one polymer to the next generally exceed those in $c_p$ itself.
Indeed, Fig.~\ref{fig:cpall} reveals that this appears to be the case. 
It shows our results for the final specific heat of polymers, for which we could not find experimental results in the temperature range, where the polymers can be equilibrated, but only at lower temperature for the experimentally and technologically relevant polymers, PAP, PAM, PAA, and PNIPAM~\cite{Cahill16Mac}.
Computed $c_p$ values together with $\Delta c_p$ estimates are listed in the Supplementary Table S1~\cite{SM}.

\subsection{United-atom simulations}

The explicit-atom model simulations were repeated for a united-atom model of hexadecane~\cite{Martin1998JPCB}.
The crucial task of estimating $g(\nu)$ is now divided into two parts: the computation of the spectra associated with the explicitly treated units and that of the missing DOFs. 
The course of action differs depending on which of the three methods proposed in Sect.~\ref{sec:methodB} to estimate $c_p$ from united-atom-based simulations is chosen.
However, in either case, the first step is  to deduce $g(\nu)$ for the united atoms.

\begin{figure*}[ptb]
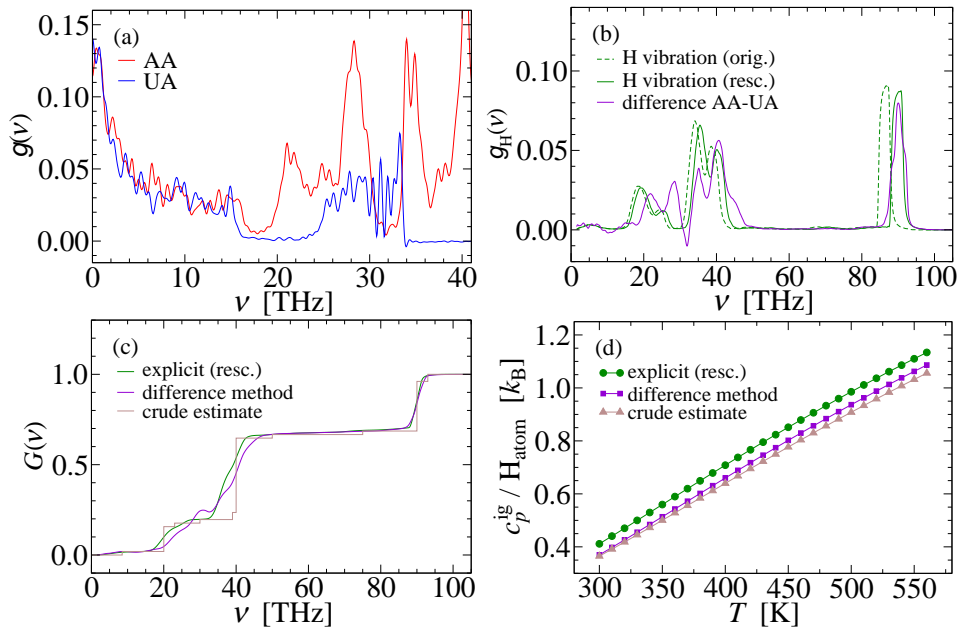

    \centering
    \hspace{-7pt}
    \includegraphics[width=.348\textwidth]{spectrum_ua_aa.eps}
    \includegraphics[width=.35\textwidth]{spectrum_H.eps}
    \includegraphics[width=.336\textwidth]{spectrum_integration.eps}
    \hspace{0pt}
    \includegraphics[width=.342\textwidth]{cp_ua_ignore.eps}
    \caption{(a) A comparison of spectra of $n$-hexadecane from all-atom and united-atom models at 430 K. (b)  
    The difference spectrum (diff), $g_\textrm{diff} \equiv g_\textrm{AA}-g_\textrm{UA}$ is compared to the explicit-H spectrum $g_\textrm{H}$ obtained as described in Sect.~\ref{sec:methodB}.
    The latter is shown in its original (orig.) and rescaled (resc.) version in green dotted and solid lines, respectively. 
    (c) Integral over the spectra shown in (b). Here, the data for the crude estimation is obtained by the weighted linear combination
    of the data shown in Fig. S3. (d) Ignored $c_p$ of $n$-hexadecane in united-atom models retrieved via the three approaches described in Sec. III.B.
    } 
    \label{fig:UA-spectra}
\end{figure*}

Fig.~\ref{fig:UA-spectra}(a) reveals that the low-frequency part of the UA and AA spectra ($\nu \lesssim 16$~THz, {related to C-C-C bond angle vibrations}) are quite similar.
The first peak missing in the UA spectrum lies slightly above $\nu = 20$~THz, which can be associated with torsional vibrations of terminal CH$_3$ groups.
The highest frequencies in the UA spectrum, i.e., those slightly above 30~THz, can be associated with united-atom bond vibrations.

The difference between all-atom and united-atom spectra (reweighted to the true number of DOFs), $g_\textrm{H}(\nu)$, is shown in Fig.~\ref{fig:UA-spectra}(b) (violet solid line) and compared to the spectrum that is obtained when all carbon atoms are frozen in and only the hydrogen atoms are explicitly propagated (green dashed line). 
Qualitative agreement is obtained, which, however, is further improved when rescaling the explicit spectrum according to $g(\alpha \nu)/\alpha$ with $\alpha = \sqrt{13/12}$ (green solid line).
The integral over $G(\nu) \equiv \int_0^\nu \! \textrm{d}\nu'\,g_\textrm{H}(\nu')$ can be approximated as a linear combination of step function, whose derivative is given in Eq.~(\ref{eq:IDW2a}), which is demonstrated in Fig.~\ref{fig:UA-spectra}(c). 
It turns out that the different methods to account for the ignored density of states does not effect strongly the predicted $\Delta c_p$.
They differ by at most 0.05~$k_\textrm{B}$ in the investigated temperature interval as demonstrated in Fig.~\ref{fig:UA-spectra}(d). 

\begin{figure}[hbtp]
    \centering
    \includegraphics[width=.4\textwidth]{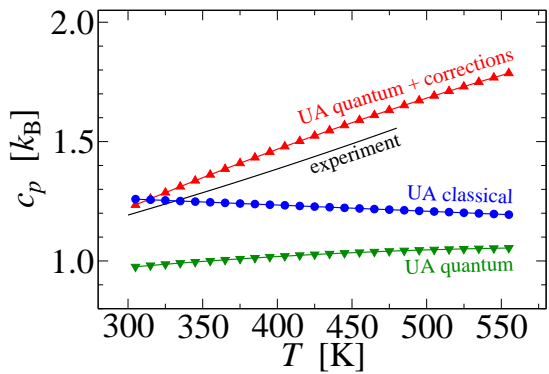}
    \caption{
    Specific heat $c_p$ of $n$-hexadecane as a function of temperature:
    experimental data (black lines), uncorrected $c_p$ of a classical, united-atoms based simulation before (blue circles) and after (green triangles down) applying quantum corrections as well as full estimates for $c_p$ (red triangles up) obtained using Eq.~(\ref{eq:fullUA-cp}), which includes corrections for ignored H~atoms. Thee experimental data on $c_p$ is taken from Ref.~\cite{Regueira2017JCT}
    }
    \label{fig:UAcp}
\end{figure}

Finally, we find that $c_p$ as predicted with a UA potential from classical simulations near room temperature might falsely be believed to be accurate, since values turn out close to experimentally measured values, see Fig.~\ref{fig:UAcp}.
However, $c_p$ (of $n$-hexadecane) decreases upon heating in UA classical simulations, while it increases experimentally.
To make accurate predictions for the right reason, the specific heat must be corrected, e.g., in one of the three ways proposed in Sect.~\ref{sec:methodB}.
This leads to an agreement within 0.1~$k_\textrm{B}$ per atom throughout the investigated temperature range with the available experimental data~\cite{Regueira2017JCT}, as revealed by Fig.~\ref{fig:UAcp}.

It is interesting to note that the united-atom potential lead again to a slight overestimation of $c_p(T)$ in comparison to the available experimental data~\cite{Regueira2017JCT}. 
This could be coincidence, however, there may also be a reason why different potentials lead to similar errors. 
Both potentials were optimized to closely match density and viscosity as a function of temperature and pressure. 
Neither one, however, includes explicitly many-body dispersion terms, which, however, are not entirely negligible for molecular systems~\cite{Elrod1994CR}.

\section{Conclusions and Outlook}
\label{sec:conclusions}

We presented a method allowing the specific heat of molecular systems to be corrected for vibrational quantum effects and demonstrated that the specific heat of various chain molecules can be computed with it so that the specific heat can be predicted as reliably from molecular simulations as any other quantity. 
In principle, the presented method also applies to systems other than chain molecules.
In fact, it will most likely improve the specific heat prediction of any classically treated system with vibrational frequencies above what we call thermal frequencies.
However, the method does not capture quantum-mechanical anharmonicity effects, as they occur in a non-negligible way, for example, in the case of water at room temperature~\cite{Tuckerman1997S}. 
Likewise, whenever the temperature of a system is below its Debye temperature, anharmonicity will effect the specific heat to some degree.  
For a truly accurate computation of the specific heat of such systems, we see no way around the use of path-integral simulations~\cite{Tuckerman1993JCP,Muser1995PRB,Martonak1998PRE}.
However, for molecules with closed valence shell other than a few small selected molecules, such as water, methane, and ammonia, any intermolecular (including rotational) motion can be classified as classical at room temperature.  

Of course, even for polymers---like the ones investigated in this study---anharmonic quantum effects do exist. 
To compute them using an all-atom framework, it may not be necessary to use Trotter numbers as large as $P \gtrsim h\nu_\textrm{max}/(k_\textrm{B} T)$, where maximum frequencies are typically associated with vibrations of terminating hydrogen atoms. 
The idea to compute a mass-weighted velocity auto-correlation function to correct for an insufficient handling of intramolecular, vibrational quantum effects, which we presented in this work, can be generalized to path-integral simulations.
This is possible, because it can be readily worked out how the predicted specific heat of a harmonic reference depends on the Trotter number $P$ and the ratio $h\nu/(k_\textrm{B} T)$ so that the excess specific heat obtained at finite $P$ can be estimated. 
Such an approach should be particularly beneficial when intermolecular interactions are clearly weaker than intramolecular forces but not necessarily for regular metals and ceramics.

An indirect result of our study is that replacing hydrogen atoms with deuterium  would not only enhance their chemical stability due to a reduction of zero-point energy, which was argued to benefit the tribological properties of hydrogen terminated coatings~\cite{Mo2009PRB}, but it would also increase the specific heat and thereby presumably the heat conduction.
We estimate the increase in $c_p$ due to full deuteration in paraffins and polyalphaolefins to be 0.25~$k_\textrm{B}$/atom at $T = 300$~K and at 0.3~$k_\textrm{B}$/atom at $T = 400$~K, which would correspond to an increase of roughly 25\% in the specific heat and potentially to a similar increase in heat conduction.
However, this insight is at best relevant for small-scale, niche applications,  given that the currently achieved production of deuterated mineral oils is in the decagram range~\cite{Klenner2020PC}.

A more immediate implication of our work is that a successful computation of thermal transport properties will necessitate a correct assessment of the specific heat \cite{Cahill16Mac}.
When simulations using accurate potentials are conducted carefully but a classically computed heat conductivity $\kappa$ is not reweighted with a similar factor to account for quantum effects as the specific heat, we would expect
$\kappa$ to be overestimated~\cite{KappaMDExp,Mukherji19PRM}.
This might explain why one of us~\cite{Mukherji19PRM} found $\kappa \simeq 0.304$~W/Km and 0.264~W/Km for \textit{in-silico} PMMA and PAP, respectively, while the corresponding experimental values are 0.200~W/Km and 0.160~W/Km \cite{Cahill16Mac}.

\section{Acknowledgement}
M.M. thanks Markus Gallei for useful discussions. 
D.M. thanks the Canada First Research
Excellence Fund (CFREF) for financial support and the ARC Sockeye computational facility where the commodity polymer simulations are performed.

\end{document}